\documentclass[universe,article,accept,pdftex,oneauthor]{Definitions/mdpi} 

\firstpage{1} 
\pubvolume{1}
\issuenum{1}
\articlenumber{0}
\pubyear{2024}
\copyrightyear{2024}
\externaleditor{Academic Editor: Firstname\\ Lastname}

\usepackage[normalem]{ulem}

\datereceived{27 April 2024} 
\daterevised{23 May 2024} 
\dateaccepted{28 May 2024} 
\datepublished{ } 
\hreflink{https://doi.org/} 
\pdfoutput=1 

\Title{{{Mechanisms}} 
 for Producing Primordial Black Holes from Inflationary Models Beyond Fine-Tuning}

\TitleCitation{Mechanisms for Producing Primordial Black Holes from Inflationary Models Beyond Fine-Tuning}

\Author{{Ioanna Stamou} 
 \orcidA{} }

\AuthorNames{Ioanna Stamou}

\AuthorCitation{{Stamou, I} 
}

\address[1]{%
{Service} 
 de Physique Th\'eorique, Universit\'e Libre de Bruxelles (ULB), Boulevard du Triomphe, CP225, \mbox{B-1050 Brussels, Belgium;} ioanna.stamou@ulb.be}

\abstract{In this study, we present an  analysis of the fine-tuning required in various inflationary models in order to explain the production of Primordial Black Holes (PBHs). We specifically examine the degree of fine-tuning necessary in two prominent single-field inflationary models: those with an inflection point and those with step-like features in the potential. Our findings indicate that models with step-like features generally require less fine-tuning compared to those with an inflection point, making them  more viable for consistent PBH production. An interesting outcome of these models is that, in addition to improved fine-tuning, they may also predict low-frequency signals that can be detected by pulsar timing array (PTA) collaborations. 
Additionally, we extend our analysis to multifield inflationary models to assess whether the integration of additional fields can further alleviate the fine-tuning demands. The study also explores the role of a spectator field and its impact on the fine-tuning process.
Our results indicate that although mechanisms involving a spectator field can circumvent the issue of fine-tuning parameters for PBH production, both multifield models and models with step-like features present promising alternatives. 
While fine-tuning involves multiple considerations, our primary objective is to evaluate various inflationary models to identify the one that most naturally explains the formation of PBHs. 
{Hence, this study introduces a novel approach by categorizing existing PBH mechanisms, paving the way for subsequent research to prioritize models that minimize the need for extensive fine-tuning.}
}

\begin{document}


\section{Introduction}

In recent years, there has been an increasing focus on primordial black holes (PBHs) due to their possible contribution to explaining the dark matter of the Universe. This explanation has gained significant attention following the detection of gravitational waves (GWs) emitted by binary black hole mergers, as reported by the LIGO/VIRGO/Kagra (LVK) collaboration \cite{Abbott:2016blz, Abbott:2016nmj, Abbott:2017vtc, Abbott:2017gyy, Abbott:2017oio}. Moreover, the recent detection of GWs at nano-Hertz frequencies by pulsar timing array (PTA) collaborations \cite{NANOGrav:2023gor, NANOGrav:2023hde, NANOGrav:2023hvm, EPTA:2023fyk, EPTA:2023sfo, EPTA:2023akd, EPTA:2023xxk} has sparked additional interest in the potential link between these low-frequency GWs and PBHs . Consequently, numerous theoretical studies have explored the potential link between dark matter, PBHs, and GWs \cite{Ozsoy:2023ryl, Sasaki:2018dmp, Carr:2023tpt, Khlopov:2008qy,Kuhnel:2024zpv,Ellis:2023oxs,Vaskonen:2020lbd,Escriva:2022duf}. All these studies must adhere to the observational constraints on inflation from Cosmic Microwave Background (CMB) anisotropies as reported by the Planck collaboration~\cite{Planck:2015sxf,Planck:2018jri}.

 In  previous theoretical research, it has been suggested that PBHs can form through a significant enhancement of curvature  perturbations at small scales. One widely adopted mechanism for producing PBHs from inflation involves studying an inflection point in the effective scalar potential \cite{Garcia-Bellido:2017mdw,Hertzberg:2017dkh,Ballesteros:2017fsr,Ballesteros:2019hus,Cicoli:2018asa,
Stamou:2021qdk,Spanos:2022euu,Dalianis:2018frf,Germani:2017bcs,Ezquiaga:2017fvi,Ozsoy:2018flq,Ozsoy:2020kat,Karam:2022nym,Aldabergenov:2023yrk,Kannike:2017bxn,Kawasaki:2016pql,Choudhury:2013woa}. Such inflationary potentials are generally characterized by a slow-roll phase at large  scales exiting the horizon
, followed by an inflection point region at small scales where the power spectrum undergoes a significant enhancement. However, a major drawback of this mechanism is the high level of fine-tuning that the parameters require to achieve such a configuration in the potential \cite{Stamou:2021qdk,Hertzberg:2017dkh,Cole:2023wyx}. 
{ It is important to note that, within this work, `fine-tuning' refers to the necessity for highly precise parameters to facilitate the abundant production of PBHs. A model is deemed fine-tuned if minor variations in parameters significantly influence the enhancement of the power spectrum and, consequently, the production of PBHs.}
 
Alternative mechanisms within the framework of single-field inflation have also been proposed to alleviate problematic fine-tuning \cite{Inomata:2021tpx,Cotner:2017tir,Kefala:2020xsx,Dalianis:2021iig,Cai:2021zsp,Choudhury:2023jlt,Choudhury:2024jlz,Gu:2023mmd,Drees:2011yz,Domenech:2023dxx,Caravano:2024tlp,Sharma:2024whg,Dimastrogiovanni:2024xvc,Karam:2023haj}. Mechanisms involving step-like features in  potentials  have been studied \cite{Inomata:2021tpx,Kefala:2020xsx,Dalianis:2021iig,Cai:2021zsp}. These mechanisms can lead to a significant enhancement of the power spectrum, which is capable of explaining the production of PBHs. Interestingly, this mechanism predicts additional oscillatory patterns in the power spectra that may  lead to a characteristic signal in the induced GWs \cite{Fumagalli:2020nvq}
.

 The study of an amplification of scalar power spectrum {has also been extended}  to multi-field models. Multi-field inflationary models have been widely explored for the production of PBHs \cite{Palma:2020ejf,Anguelova:2020nzl,Iacconi:2023slv,Zhou:2020kkf,Mavromatos:2022yql,Kawai:2022emp,Aldabergenov:2022rfc,Braglia:2020eai,Pi:2021dft,Gundhi:2020kzm,Geller:2022nkr,Chen:2023lou,Wang:2024vfv}. Models with a non-canonical kinetic term in the Lagrangian can lead to a significant enhancement in the scalar power spectrum \cite{Aldabergenov:2022rfc,Braglia:2020eai,Pi:2021dft,Gundhi:2020kzm,Geller:2022nkr,Chen:2023lou,Wang:2024vfv,Heydari:2021gea,Pi:2017gih}. Moreover, multi-field models within the framework of natural inflation, such as those involving multiple axion fields \cite{Zhou:2020kkf, Mavromatos:2022yql}, have been shown to result in a significant fractional abundance of PBHs.
 Finally,  hybrid models with a waterfall trajectory have acquired a lot of interest as they provide a more natural explanation to PBHs \cite{Afzal:2024xci,Spanos:2021hpk,Braglia:2022phb,Clesse:2015wea,Tada:2023fvd,Dimopoulos:2022mce}.

Furthermore, mechanisms involving a spectator field have been extensively explored as promising alternatives to mitigate the fine-tuning required for PBH production \cite{Carr:2017edp,Kohri:2012yw,Stamou:2023vft,Stamou:2023vwz,Stamou:2024xkk,Cai:2021wzd,Gow:2023zzp,Hooper:2023nnl}. Although the spectator field does not drive inflation, it evolves independently and significantly influences the formation of structures and phenomena, including PBHs. Notably, it has been demonstrated that these models can circumvent the need for fine-tuning the potential's parameters to facilitate PBH production \cite{Kohri:2012yw,Stamou:2023vft}.

{  This paper investigates the extent of fine-tuning required across a diverse range of mechanisms previously proposed in the literature. Specifically, we examine four characteristic inflationary mechanisms for producing primordial black holes (PBHs): a single-field mechanism with an inflection point, one with a step-like feature, a two-field mechanism with a non-canonical kinetic term, and a mechanism incorporating a waterfall trajectory. For each PBH production mechanism, we provide a rationale for selecting a particular model. This study diverges from the findings reported in Ref.~\cite{Cole:2023wyx} as it does not solely concentrate on single-field inflationary models with an inflection point or bulk in their scalar potential. While acknowledging the contributions of these mechanisms, our analysis broadens to include additional inflationary models, such as those involving multiple fields, as discussed earlier.}


 First, we delve into single-field inflation models, examining both the inflection point mechanism and models that feature step-like characteristics while evaluating the associated fine-tuning demands. We found that the amount of fine-tuning required for models with an inflection point is significantly large, whereas the step-like mechanism can lead to a more natural explanation as the amount of fine-tuning is significantly reduced. Notably, this mechanism can explain GW signals at low frequencies. We discuss this implementation in a separated section and leave the details for a future paper.

In the continuation of our analysis, we extend our study to {  analyze the fine-tuning of} multifield inflationary models. Specifically, we explore two-field model mechanisms: the first incorporates non-canonical kinetic terms in the Lagrangian, and the second features a waterfall trajectory typical of hybrid inflationary models. We find that the amount of fine-tuning is significantly reduced in both multifield mechanisms by at least three orders of magnitude in comparison with inflection point models. Lastly, we discuss the implications of models that include a spectator field for explaining the production of PBHs, highlighting that the need for fine-tuning to describe such structures can be avoided. {Therefore, the novelty of this work lies in classifying the proposed PBH mechanisms, focusing on determining which mechanisms can more naturally explain the production of PBHs. This approach can contribute to future investigations towards the mechanisms that inherently require less fine-tuning.   }

The layout of this paper is  as follows: In Section \ref{sec:fine-tuning}, we describe our methodology for computing the amount of fine-tuning. In Section \ref{sec:single_field}, we present an analysis of fine-tuning in single-field inflationary mechanisms. In Section \ref{sec:two_field}, we extend this analysis to two-field models. In Section \ref{sec:two_spectator}, we discuss the amount of fine-tuning in models with a spectator field. In Section \ref{sec:pta}, we explore an interesting application of step-like potentials to scalar-induced GWs. Finally, in Section \ref{sec:con}, we present our conclusions.  {The Appendices provide the detailed equations that are needed for our analysis: Appendix  \ref{app:a} covers the evaluation of the scalar power spectrum, Appendix \ref{app:b} details the fractional abundances of PBHs, and Appendix \ref{app:c} discusses the energy densities of GWs.}



\section{Fine-Tuning}
\label{sec:fine-tuning}

In this study, we focus on evaluating the degree of fine-tuning required by various mechanisms to explain the production of PBHs. It is important to note that there are several aspects of fine-tuning that need consideration. However, our primary goal is to assess different inflationary mechanisms to determine which can provide a more natural explanation for PBH formation.

To assess the fine-tuning of parameters, we employ the quantity $\Delta_{p_i}$, defined as the maximum value of the following logarithmic derivative \cite{Barbieri:1987fn}:
\begin{equation}
\Delta_{p_i}^Q = \mathrm{Max} \left| \frac{\partial \ln(Q) }{\partial \ln(p_i) } \right|
\label{eq:delta}
\end{equation}

Here, $Q$ represents the quantity under study, specifically the peak value of the scalar power spectrum denoted as $P_R^{peak}$, and $p_i$ denotes the corresponding parameters of each model being analyzed. This metric, $\Delta_{p_i}^Q$, quantifies the sensitivity of the observable $Q$ to variations in parameter $p_i$, thus serving as a measure of the required fine-tuning. A high value of $\Delta_{p_i}^Q$ implies that minor changes in $p_i$ result in significant alterations in $Q$, indicating a high degree of fine-tuning within the model.

We need to remark here that throughout this work, we use reduced Planck mass units, setting $\rm{M_P=1}$.


\section{Single-Field Models}
\label{sec:single_field}
In this section, we discuss single-field inflationary models for the production of PBHs. These models are widely recognized in the literature for their role in the formation of PBHs and offer several advantages. First, they provide a simple mechanism for producing PBHs during inflation. Second, they can explain   a significant fraction of dark matter. Third, they are easily adaptable to various theoretical frameworks.

In the subsequent sections, we will explore two distinct mechanisms. The first involves a mechanism characterized by an inflection point in the scalar potential \cite{Garcia-Bellido:2017mdw,Hertzberg:2017dkh,Ballesteros:2017fsr,Ballesteros:2019hus,Cicoli:2018asa,
Stamou:2021qdk,Spanos:2022euu,Dalianis:2018frf,Germani:2017bcs,Ezquiaga:2017fvi,Ozsoy:2018flq,Ozsoy:2020kat,Karam:2022nym,Aldabergenov:2023yrk,Kannike:2017bxn}. The second examines models that incorporate step-like features \cite{Kefala:2020xsx,Dalianis:2021iig}. We compute the amount of fine-tuning for both of these two different mechanisms. Our  results suggest that models with step-like features can significantly  reduce this required tuning of the parameters. 

\subsection{Inflection Point}
\subsubsection{Description of  the Mechanism}

Significant peaks in the scalar power spectrum, which are indicative of the production of PBHs and GWs, can be generated by a near inflection point in the scalar potential. This inflection point is characterized by the conditions:
\begin{equation}
\begin{split}
&\frac{dV(\phi_{\text{inf}})}{d\phi_{\text{inf}}} \approx 0, \quad \frac{d^2V(\phi_{\text{inf}})}{d\phi_{\text{inf}}^2} = 0,
\end{split}
\label{eq:cond_inflection_point}
\end{equation}
where $\phi_{\text{inf}}$ denotes the inflection point. At this point, the velocity of the inflaton field significantly decreases, leading to a local enhancement while the inflaton field remains almost constant. During this plateau, the power spectrum experiences significant enhancement, facilitating the production of PBHs.   
In this work, we also include in this category models with bump features in the potential, such as those in \cite{Mishra:2019pzq,Ozsoy:2020kat,Fu:2022ssq}, since they satisfy the conditions given in Equation~(\ref{eq:cond_inflection_point}).

As we mentioned before, the concept of an inflection point in the potential has been widely adopted in theoretical studies. The degree of fine-tuning required for inflection point models has been extensively analyzed, notably in Ref. \cite{Cole:2023wyx}. According to this study, polynomial potentials appear to require substantial fine-tuning both to explain the formation of PBHs and to meet the observational constraints of inflation at CMB scales.

\subsubsection{the Issue of Fine-Tuning}
Among the various models proposing an inflection point, we focus on the one detailed in Ref. \cite{Garcia-Bellido:2017mdw}. This model is also highlighted in Ref. \cite{Cole:2023wyx} as one of the models requiring minimal fine-tuning within the inflection point framework. The potential, as described in Ref. \cite{Garcia-Bellido:2017mdw}, is given by:
\begin{equation}
V(\phi)=\frac{\lambda}{12} \phi^2 \upsilon^2 \left(\frac{6 - 4a \frac{\phi}{\upsilon} + 3 \frac{\phi^2}{\upsilon^2}}{\left(1 + b \frac{\phi^2}{\upsilon^2}\right)^2}\right)
\label{eq:potential_inf}
\end{equation}

Here, $a$, $b$, $\lambda$, and $\upsilon$ are the parameters of the model. The parameter $\lambda$ sets the scale of the power spectrum at CMB scales, while the others primarily determine the shape of the potential. A significant enhancement of the power spectrum, capable of facilitating the production of PBHs, can be achieved with the following specific selection of parameters:
\begin{equation}
a = 1/\sqrt{2}, \quad b ={ 3/2}, \quad \lambda =1.86\times 10^{-6} , \quad \upsilon = 0.19669
\label{eq:parameters_slow_roll}  
\end{equation}
{and} 
 for initial condition we assume $\phi_{ic}=2.4719$ ({{we assume always that the initial conditions are taken at the pivot  scales ($\rm{k=0.05 M_{pc}^{-1}}$) in order to make a prompt comparison with the observable constraints of inflation}}). This choice of initial condition led to the following prediction of the spectral index, $\rm{n_s}$, and the tensor-to-scalar ration, $\rm{r}$ ({{for the evaluation of spectral index} $n_s$ and $r$, we assume that $n_s  \simeq 1+2 \eta_V -6\varepsilon_V$ and $r \simeq 16 \varepsilon_V$, where $ \varepsilon_V= \frac{1}{2}\left({V'(\phi)}/{V(\phi)} \right)^2 , \quad \eta_V={V''(\phi)}/{V(\phi)}$.}): 

\begin{equation}
n_s=0.956 ,\quad r= 0.0064 \,. 
\end{equation}

\begin{figure}[h!]
\includegraphics[width=10.5 cm]{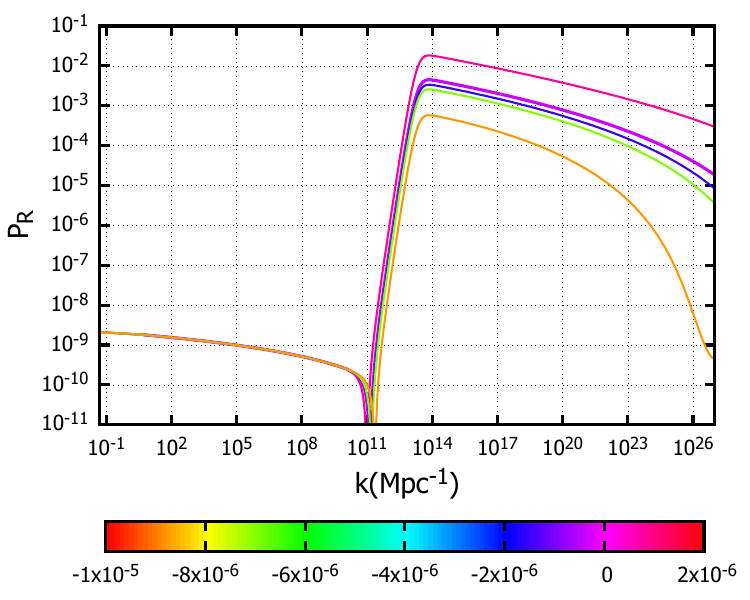}
\caption{{The power} 
 spectrum derived from the potential described in Equation~(\ref{eq:potential_inf}), showing the effects of small variations in the parameter b of order $\mathcal{O}(10^{-6})$. {The purple line corresponds to the parameters given in Equation~(\ref{eq:parameters_slow_roll}).}
\label{fig1}}
\end{figure}

 In Figure~\ref{fig1}, \textls[-15]{we depict the power spectrum for a small variation of the parameter \(b\), which influences the inflection point and consequently requires more fine-tuning. This figure clearly demonstrates that the peak value is highly sensitive to variations in this~parameter.}

To systematically quantify the fine-tuning, we evaluate the quantity of Equation~(\ref{eq:delta}). ({{The evalution of power spectrum is briefly described in Appendix }\ref{app:a}.)} In Table \ref{tab1}, we present the evaluation of Equation~(\ref{eq:delta}) for the potential (\ref{eq:potential_inf}). Our results are consistent with those proposed by Cole et al. \cite{Cole:2023wyx} in their study of fine-tuning for the peak of the scalar power spectrum. The parameter $\rm b$ can be given by the parameter $\rm a$ by the following expression~\cite{Garcia-Bellido:2017mdw,Cole:2023wyx}:
\begin{equation}
b = 1-\frac{a^2}{3}+\frac{a^2}{3}\left( \frac{9}{2a^2} -1 \right)^{2/3}\, .
\end{equation}

This parameter defines the inflection point. In contrast to \cite{Cole:2023wyx}, we also evaluate the fine-tuning of the parameter {$b$} 
 individually, raising questions about the necessity of redefining this function and uncovering hidden fine-tuning, and we found that this parameter need a lot of fine-tuning. As discussed in \cite{Cole:2023wyx}, the amount of fine-tuning required for models with an inflection point is significant; with variations in the studied models, it can reach an order of magnitude as high as $\rm {10^{8}}$.
The fine-tuning associated with models featuring an inflection point has also been studied in \cite{Stamou:2021qdk, Spanos:2022euu}.

\begin{table}[H] 
\caption{Analysis of fine-tuning for the potential given in Equation~(\ref{eq:potential_inf}).\label{tab1}}
\begin{tabularx}{\textwidth}{CC}
\toprule
\textbf{Parameter}	& \boldmath{$\Delta_{p_i}^{P_R^{peak}}$}	\\
\midrule
a		& $\rm{7.1 \times 10^{2}}$				\\
$\upsilon$		& $\rm{4.6 \times 10^{2}}$			\\
b		& $\rm{4.2 \times 10^{5}}$ 			\\
\bottomrule
\end{tabularx}
\end{table}

In our analysis, we also evaluate the degree of fine-tuning required for the fractional abundance of PBHs, denoted as $f\rm{_{PBH}}$. So, we compute the $\Delta_{p_i}^{f_{\rm{PBH}}}$  from Equation~(\ref{eq:delta}). We found that 
\begin{equation}
{\Delta_{p_i}^{f_{\rm{PBH}}}}/{\Delta_{p_i}^{P_R^{peak}}} \approx 30
\end{equation}
which indicates that achieving the desired fractional abundance necessitates fine-tuning that is {1--2 orders of magnitude} more stringent than that required for the parameters outlined in Table \ref{tab1}. This elevated level of fine-tuning underscores the sensitivity of PBH formation to the specific dynamics of the inflationary model parameters. { The evaluation of the fractional abundance of PBHs,  $f_{\text{PBH}}$, is detailed in Appendix \ref{app:b}. In this Appendix, we also present the fractional abundances of PBHs for all the mechanisms discussed in this~work.
 }


\subsection{Step-Like Potential}

 
\subsubsection{An Alternative Single-Field Mechanism}
In the pursuit of models that require minimal fine-tuning, the step-like potential emerges as a promising approach in single-field inflationary theories \cite{Kefala:2020xsx,Dalianis:2021iig}. This potential reads as:
\begin{equation}
V(\phi) = V_0\left( 1+ \frac{1}{2}\sum_{i} A_i \left(1+\tanh\left[c_i\left(\phi-\varphi_i \right) \right]\right) +B\phi \right) \, ,
\label{eq:steppote}
\end{equation}
where $V_0$  is fixed for the requirement of the amplitude of power spectrum at CMB scales; {$\phi$ is the inflaton field};  and $A_i$, {$B$}, $c_i$, and $\varphi_i$ are parameters that define the amplitude, steepness, and position of the steps, respectively. These steps introduce localized features in the potential, enabling significant enhancements in the scalar power spectrum.

A step-like feature in the inflaton potential results in a rapid decrease in potential energy and leads to spectral distortions \cite{Kefala:2020xsx,Dalianis:2021iig,Dalianis:2023pur}. Such features, where the potential energy abruptly transitions from one constant value to another at one or more points, can significantly enhance the spectrum over a specific range of scales.

{The distinctive behavior of step-like features in the potential contrasts markedly with other inflationary mechanisms, such as the inflection point mentioned previously. While the inflection point leads to a temporary stop in the field's motion, leading to a spike in the power spectrum due to prolonged inflation at specific field values, a step-like feature triggers accelerations and decelerations in the inflaton's velocity. For more details on the differences between these mechanisms, see Appendix \ref{app:a}. Unlike the inflection point, the enhancement in the power spectrum in a step-like potential does not rely on the minimum value of the field's velocity. For this reason, we anticipate that this mechanism can reduce the amount of fine-tuning required.}

\subsubsection{Studying the Fine-Tuning}

The potential described in Equation~(\ref{eq:steppote}) can be integrated into the $\alpha$-attractor framework of supergravity,
{ providing a reliable model that adheres to the observable constraints of the CMB, and can lead to an enhancement in the power spectrum \cite{Dalianis:2021iig}. However, this model does not achieve a significant enough enhancement of the power spectrum for the production of PBHs. For this reason, in our study focusing on assessing fine-tuning, we modified the well-known Starobinsky potential to incorporate step-like features. }{Hence, in our study, we have adopted the following form of the potential that aligns with the observable constraints:}
\begin{equation}
V (\phi) = (1 - e^{-d \phi})^2 \left( V_0 + \sum_{i=0}^N B_i \left(-1 + \tanh[c (\phi - \varphi_i)]\right) \right)
\label{eq:steppot}
\end{equation}
where $\rm{V_0}$, $\rm{c}$, $\rm{d}$,  and ${\varphi_i}$ are parameters. The parameters $\rm{V_0}$ and $\rm{d}$ ensure that the potential with respect to the CMB constraints of inflation and the others identifies the steepness and the position of the step as previously
.

In this study, we focus on the potential given in Equation~(\ref{eq:steppot}) to explore a substantial enhancement of the scalar power spectrum. The following parameters are proposed to achieve such an enhancement at $\mathcal{O}(10^{-2})$ orders of magnitude:
\begin{equation}
\begin{split}
V_0 =&4 \times 10^{-9} ,\quad B_i=0.05(V_0/2), \quad d=0.76 \quad c=1000\\
\varphi_0=&0,\quad \varphi_1=5,\quad \varphi_2=5.2,\quad \varphi_3=5.4,\quad \varphi_4=5.6\quad \, .
\end{split}
\end{equation}

For the initial condition, we assume $\phi_{ic}=6$. The prediction for the spectral index, $\rm{n_s}$, and the tensor to scalar ratio, $\rm{r}$, are given, respectively, as:
\begin{equation}
n_s=0.971 ,\quad r= 0.0026\, 
\end{equation}
{and} they  respect the observable constraints of inflation \cite{Planck:2018jri}. 

Potential (\ref{eq:steppot}) includes the four steps necessary to achieve the desired enhancement level. The dependence of the power spectrum's peak height on the parameter $c$ is depicted in Figure~\ref{fig2}. The selection of this potential facilitates the formation of PBHs at \(10^{-6} M_{\odot}\). { In Appendix \ref{app:b}, there is a detailed analysis for this mass value and the corresponding fractional abundances of PBH}. Different outcomes can be achieved by varying the choice of the parameter. A proper choice of parameters that can lead to significant results is presented in a subsequent section.

\begin{figure}[H]
\includegraphics[width=10.5 cm]{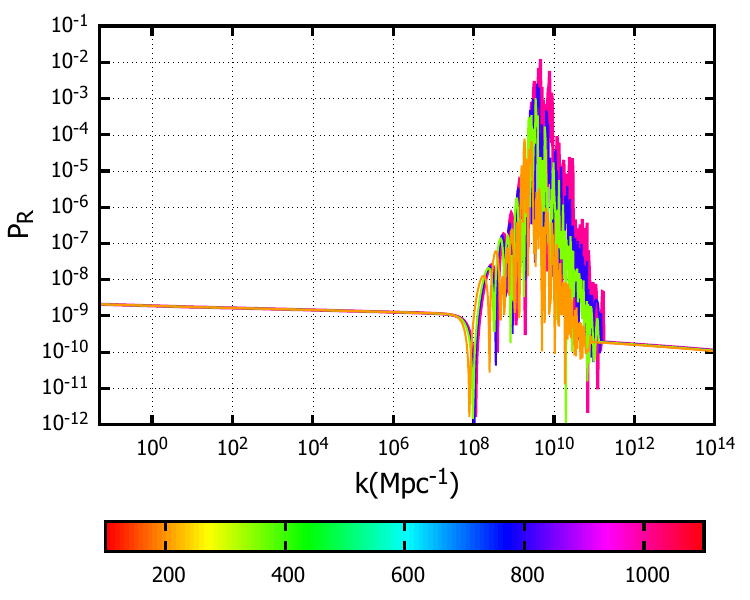}
\caption{{The} 
 power spectrum derived from the potential described in Equation~(\ref{eq:steppot}) for various choices of the parameter c.
\label{fig2}}
\end{figure}

 As previously described, we evaluate the parameter $\Delta_{p_i}^{P_R^{peak}}$, defined in Equation~(\ref{eq:delta}), and present our findings in Table \ref{tab2}. We study the parameters ${B_i}$,  ${c}$, and ${d}$,  which are responsible for the step features. We also examine the distance of the steps as $|\phi_{i+2}-\phi_{i+1}|$, where $\rm{ i=0,1}$, etc. As we can notice by Table \ref{tab2}, the parameter that needs more fine-tuning in this case is the distances of the steps $|\varphi_{i+2}-\varphi_{i+1}|$.
  However, it is noteworthy that this mechanism exhibits a substantial reduction  in the amount of fine-tuning compared to the earlier  case of an inflection point in the scalar potential, see Tables \ref{tab1} and \ref{tab2}.
\begin{table}[H] 
\caption{Analysis of fine-tuning for the potential given in Equation~(\ref{eq:steppot}).\label{tab2}}
\begin{tabularx}{\textwidth}{CC}
\toprule
\textbf{Parameter}	& \boldmath{$\Delta_{p_i}^{P_R^{peak}}$}	\\
\midrule
$B_i/V_0$		& 18.2			\\
$c$		& 3.10	\\
$d$		& 53.5		\\
$|\varphi_{i+2}-\varphi_{i+1}|$		& $80.6$		\\
\bottomrule
\end{tabularx}
\end{table} 

{ As we observe in this section, the distinct dynamics of single-field mechanisms—such as the inflection point and step-like features—lead to localized enhancements in the power spectrum. Models with step-like features in their potentials are characterized by sharper peaks and additional oscillatory patterns, unlike the smoother power spectra seen in other potential forms, such as the previously discussed inflection point. Such localized enhancements are crucial for producing PBHs within narrower mass ranges, thus influencing their  distinct characteristics (see Appendix \ref{app:b}) }

{Finally, it is crucial to acknowledge that both of the single-field mechanisms discussed have faced scrutiny regarding one-loop corrections to the evaluation of the power spectrum, as highlighted in studies such as Refs. \cite{Kristiano:2022maq,Fumagalli:2023loc,Choudhury:2023vuj,Choudhury:2023hvf,Choudhury:2024ybk,Firouzjahi:2023aum,Inomata:2022yte}. Consequently, exploring multi-field inflationary models could not only reduce the amount of fine-tuning but also offer a viable alternative for the study of PBHs.}


\section{Multi-Field Models}
\label{sec:two_field}

Multi-field inflationary models have been extensively studied for their role in generating PBHs. {In the context of multi-field mechanisms, it is anticipated that the degree of fine-tuning required will be lessened. This is expected because the constraints imposed by the CMB power spectrum and the demands for PBH production may not be fully addressed by a single-field. However, within multi-field inflation, various mechanisms may lead to differing degrees of fine-tuning, necessitating further investigation to determine their efficacy. } 
In this section, we discuss the amount of fine-tuning required in multi-field scenarios, focusing on two distinct mechanisms: one with a non-canonical kinetic term \cite{Aldabergenov:2022rfc,Braglia:2020eai,Pi:2021dft,Gundhi:2020kzm,Geller:2022nkr,Chen:2023lou,Wang:2024vfv} and another with a canonical kinetic term accompanied by a waterfall trajectory~\cite{Afzal:2024xci,Spanos:2021hpk,Braglia:2022phb,Clesse:2015wea,Tada:2023fvd,Dimopoulos:2022mce}.    


\subsection{Two Field Model with a Non-Canonical Kinetic Term}


{In this subsection, we introduce a two-field model incorporating a non-canonical kinetic term. The enhancement of the power spectrum in this model arises through a large non-canonical kinetic coupling within a two-field inflationary framework. This mechanism is characterized by a temporary tachyonic instability of the isocurvature perturbations during the transition between two distinct inflationary phases~\cite{Aldabergenov:2022rfc,Braglia:2020eai,Gundhi:2020kzm,Geller:2022nkr}. Such instability leads to a significant amplification of the curvature perturbations, which can be imprinted in both the production of PBHs and GWs .}

{ Although numerous models with a non-canonical kinetic term have been proposed in the literature, }we specifically focus on the one detailed in Ref.~\cite{Braglia:2020eai}. The action for this two-field toy model is expressed as:
\begin{equation}
S = \int d^4 x \sqrt{-g} \left[ \frac{M_P^2}{2} R - \frac{1}{2} (\partial \phi)^2 - e^{-b_1 \phi} (\partial \chi)^2 + V(\phi, \chi) \right]\,.
\label{eq:two_field_action}
\end{equation}

In this model, the inflaton field $\phi$ interacts with a non-canonical scalar field $\chi$ via an exponential dilaton-like coupling \cite{Kallosh:2022vha}. 
The potential is defined by:
\begin{equation}
V(\phi, \chi) = V_0 \frac{\phi^2}{\phi_0^2 + \phi^2} + \frac{m_{\chi}^2}{2} \chi^2
\label{eq:two_field_potentia}
\end{equation}

Here,
$b_1$ is a coupling parameter between the fields and $V_0$, $\phi_0$, and $m_{\chi}$, which are the model parameters. We adopt the parameter choices from \cite{Braglia:2020eai} as follows:
\begin{equation}
 V_0/(m_\chi M_P)^2 = 500 \quad \text{and} \quad  \phi_0 = \sqrt{6} M_P.
\end{equation}

For the initial conditions, we have:
\begin{equation}
\phi_{ic}=7,\quad \text{and} \quad \chi_{ic}=6.55 
\end{equation}
{as}  they lead to acceptable values for the prediction of the spectral index,  $\rm{n_s}$, and the tensor to scalar ratio, r \cite{Braglia:2020eai,Planck:2018jri}.

\begin{figure}[H]
\includegraphics[width=10.5 cm]{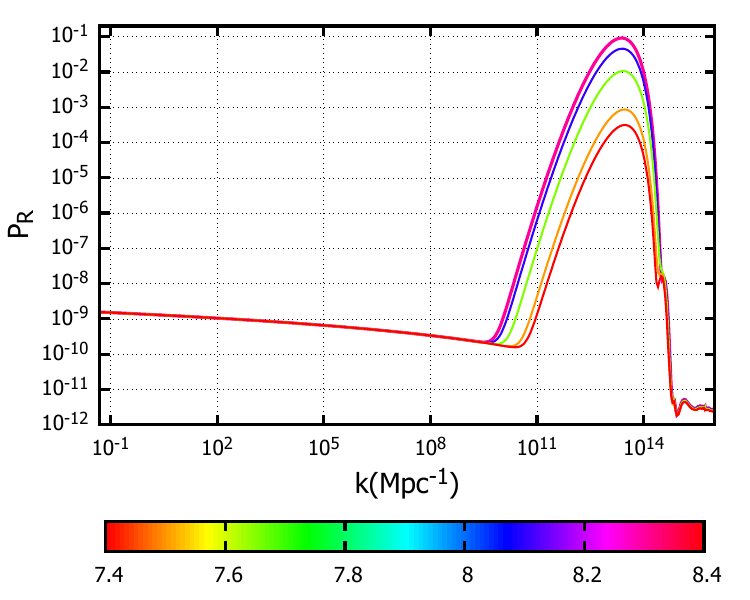}
\caption{{The} 
 power spectrum derived from Equations~(\ref{eq:two_field_potentia})  and (\ref{eq:two_field_action}) for various choices of $b_1$ present in the kinetic term
 . 
\label{fig3}}
\end{figure}

 In Figure~\ref{fig3}, we depict the power spectrum for various settings of the parameter $b_1$. It is evident that the enhancement of the power spectrum varies with the choice of this parameter. However, the impact of other parameters on the power spectrum also requires examination. Therefore, in Table \ref{tab3}, we present the degree of fine-tuning by evaluating Equation~(\ref{eq:delta}) for all parameters, including the potential parameters $m_{\chi}$ and $\phi_0$. It should be noted that the parameter $V_0$ does not influence the dynamics of the fields, and for this reason, it is not included in this table.

\begin{table}[H] 
\caption{Analysis of fine-tuning for the potential given in   Equations~(\ref{eq:two_field_potentia})  and (\ref{eq:two_field_action}).\label{tab3}}
\begin{tabularx}{\textwidth}{CC}
\toprule
\textbf{Parameter}	& \boldmath{$\Delta_{p_i}^{P_R^{peak}}$}	\\
\midrule
$m_{\chi}$		& 1.72			\\
$b_1$		& 71.7		\\
$\phi_0$		& 3.02 	\\
\bottomrule
\end{tabularx}
\end{table}  
 
The results presented in Table \ref{tab3} indicate that fine-tuning in multi-field inflationary models can be significantly reduced compared to single-field inflationary models, particularly those with an inflection point. For a more detailed analysis, it is also essential to examine how the initial conditions of the fields may influence these results.

\subsection{Hybrid Model}



{In this section, we delve into a hybrid model characterized by a waterfall trajectory. The key aspect of this mechanism is the tachyonic instability that occurs during the transition between two stages of inflation. This instability results in a significant enhancement of the power spectrum at scales corresponding to the transition. Unlike the previous mechanism where the instability arises from strong kinetic coupling, here the instability is triggered when the inflaton field reaches a critical point, leading the so-called waterfall field to develop a tachyonic solution. In contrast to the previous mechanism requiring intricate parameter setups, this hybrid model naturally derives its instability solely from the field's potential, enhancing the power spectrum in a more  predictable manner. This  approach to PBH production was first proposed in \cite{Clesse:2015wea}.}

{Before presenting a specific model for this mechanism,} it is important to outline some fundamental aspects of this framework. The hybrid model is derived from a globally supersymmetric (SUSY) renormalizable superpotential \cite{Copeland:1994vg,Dvali:1994ms}:
\begin{equation}
W = \kappa S (\bar{\Psi}_1 \Psi_2 - m^2).
\label{spotential}
\end{equation}

The $F$-term SUSY potential is given by:
\begin{equation}
V_F^{\text{SUSY}} = \kappa^2 \left[(\psi^2 - m^2)^2 + \phi^2 \psi^2 \right],
\label{firstsusy}
\end{equation}
where we have set $|S| = \phi / \sqrt{2}$ and $|\Psi_1| = |\bar{\Psi}_2| = \psi$. The field $\psi$ develops tachyonic solutions under the condition:
\begin{center}
$\kappa^2(-2m^2 + \phi^2 + 6\psi^2) < 0.$
\end{center}

Along the flat direction ($\psi = 0$), the condition becomes:
\begin{center}
$\phi^2 < \phi_c^2 = 2m^2 \equiv M^2,$
\end{center}
where $\phi_c$ represents the critical value of the field $\phi$, beyond which the field $\psi$ becomes tachyonic. The field $\psi$ is regarded as the waterfall field and the field $\phi$ as the inflaton.

A variety of hybrid models have been explored within the literature to elucidate the production of PBHs \cite{Afzal:2024xci,Spanos:2021hpk,Braglia:2022phb,Clesse:2015wea,Tada:2023fvd,Dimopoulos:2022mce}. We study  the fine-tuning aspects of hybrid models, with a particular focus on the approach outlined in Ref. \cite{Braglia:2022phb}. We selected this model due to the unique characteristics of its two-field potential, which not only serves as an attractor for the initial conditions of both fields but also ensures that, regardless of these starting conditions, the trajectory consistently leads to the potential’s plateau. Such convergence is crucial as it initiates a slow-roll phase, an important element in the dynamics of the model.
The potential is given as follows: 
\begin{equation}
        V(\phi,\psi)  =  M^2 \left[  \frac{(\psi^2-\psi_0^2)^2}{4\psi_0^2} +\frac{m^2}{2}\phi^2 +\frac{g}{2}\phi^2\psi^2+d\psi\right],
        \label{eq:hybrid}
\end{equation}
where $M$, $m$, $g$, $d$, and $\psi_0$ are parameters.  For the parameters we have the following:
\begin{equation}
M=1.47\times 10^{-5},\quad m=0.3,\quad  g=0.8,\quad \psi_0= 2.5,\quad  d=-5\times 10^{-6}
\end{equation}
{and} for initial conditions
\begin{equation}
\phi_{ic}=3.4,\quad \psi_{ic}=0\, .
\end{equation}

Hence, this model aligns with the observable constraints of inflation \cite{Planck:2018jri,Braglia:2022phb}. 

\begin{figure}[H]\vspace{-6pt}
\includegraphics[width=10.5 cm]{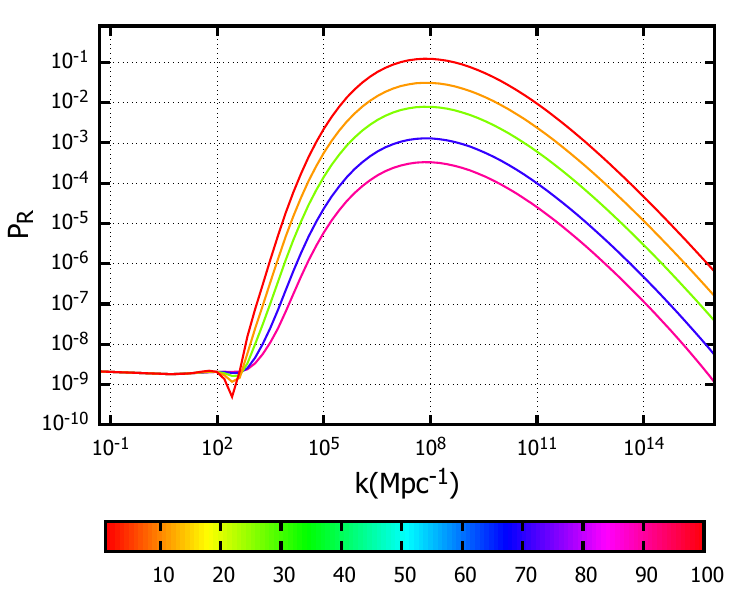}
\caption{\textls[-15]{{The} 
 power spectrum derived from Equation~(\ref{eq:hybrid}) for various choices of }parameter $\rm{10^6 |d|}$ .
\label{fig4}}
\end{figure}

In Figure~\ref{fig4}, we depict the scalar power spectrum derived from the potential in \mbox{Equation~(\ref{eq:hybrid})} for various settings of the parameter $d$. This plot illustrates that extensive fine-tuning is not necessary to achieve a significant peak in the power spectrum. As in previous mechanisms of PBHs, we analyze the impact of other parameters on the power spectrum. The parameters $m$ and $g$ influence the position of the power spectrum peak. We assess the amount of fine-tuning using Equation~(\ref{eq:delta}) and present our results in Table \ref{tab4}.

\begin{table}[H] 
\caption{Analysis of fine-tuning for the potential given in Equation~(\ref{eq:hybrid}{)} 
.\label{tab4}}
\begin{tabularx}{\textwidth}{CC}
\toprule
\textbf{Parameter}	& \boldmath{$\Delta_{p_i}^{P_R^{peak}}$}	\\
\midrule
$m$		& 5.2			\\
$g$		& 0.6		\\
$d$		& 1.9	\\
$\psi_0$		& 9.5	\\
\bottomrule
\end{tabularx}
\end{table}

As one can notice  from Table \ref{tab4}, this mechanism significantly reduces the amount of fine-tuning required, compared to both the single-field inflationary mechanism and the two-field mechanism with a non-canonical kinetic term. An important notice here is that this mechanism is not dependent on the initial condition of the fields; hence, it can provide us with  a more natural explanation of PBH production. 


\section{Spectator Field}
\label{sec:two_spectator}

From our previous analysis, it is evident that multi-field inflationary models can significantly reduce the fine-tuning required for the production of PBHs. Additionally, the implementation of a spectator field as a solution to this issue has been extensively explored in the literature \cite{Kohri:2012yw,Stamou:2023vft,Stamou:2023vwz,Stamou:2024xkk}, offering a promising way to circumvent the problem.

We specifically examine the model presented in Ref.~\cite{Stamou:2023vft},  where the spectator field experiences quantum stochastic fluctuations. These fluctuations lead the field to assume distinct mean values across different Hubble patches. Considering the vast number of these patches, it is inevitable that some will exhibit conditions allowing the spectator field to attain values necessary for quantum fluctuations to generate significant curvature fluctuations. 
  The potential of the spectator field is described by:
\begin{equation}
V(\psi) = \Lambda^4 \left( 1 - \exp\left[-\frac{\psi}{M}\right] \right)
\label{eq:pot1}
\end{equation}
where $\Lambda$ and $M$ are parameters. 

In this work, we further evaluate  the amount of fine-tuning previously assessed for models with a spectator field
. We employ the fine-tuning measure defined in Equation~(\ref{eq:delta}) as we did before. This particular mechanism does not require an enhancement in the scalar power spectrum, thereby eliminating the need to consider peaks in our evaluation. Instead, we focus on the fractional abundance of PBHs ($f_{\rm PBH}$) as our primary quantity $Q$. We specifically analyze the parameter $M$ as the parameter $\Lambda$ merely sets the scale of the spectator field and does not influence the dynamics. We found that $\Delta_{M}^{f_{\rm{PBH}}} \sim \mathcal O(1)$. Therefore, in this mechanism we can avoid the need for fine-tuning in order to explain the production of PBHs.

We need to remark here that  the stochastic nature of the spectator field during inflation plays a crucial role in suppressing fine-tuning. In this mechanism, variations in the model's single parameter $\rm M$, which is of the same order as the Hubble parameter at CMB scales, can be effectively compensated for by the specific inflationary model we study each time. A detailed analysis can be found in  \cite{Stamou:2023vft}.

\section{Step-Features Potential over Inflection-Point: an Additional Advantage in Implementation of Gws}
\label{sec:pta}

In this section, we explore the implementation of a step-like feature potential for the production of PBHs.
We demonstrate how this mechanism can not only lead to a significant enhancement and thus to the copious production of PBHs but also potentially explain the signal of GWs at low frequencies.

\begin{figure}[H]
\includegraphics[width=10.5 cm]{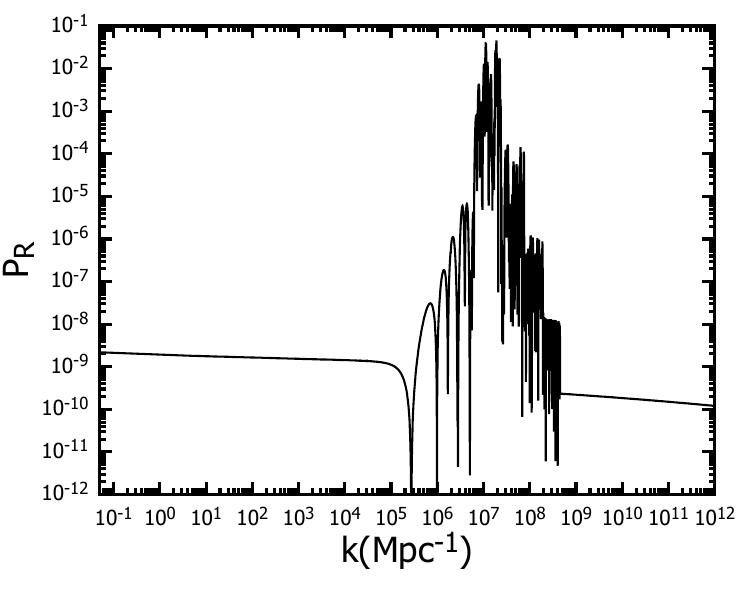}
\caption{{The} 
 power spectrum derived from the  potential given in Equation~(\ref{eq:steppot}), with the choice of parameters given in (\ref{eq:parameters_step2}).
\label{fig5}}
\end{figure}

To begin with, this approach facilitates a significant enhancement of the power spectrum while substantially reducing the need for fine-tuning. \textcolor {black}{ In this section, we consider the potential  given in Equation~(\ref{eq:steppot}) }. The reinforcement of the power spectrum  is depicted in Figure~\ref{fig5}. 
For this plot, we assume the following set of parameters:
\begin{equation}
\begin{split}
c=3000,\quad d=0.73,\quad B_i/(V_0/2)=0.049, \\   |\phi_{i+2}-\phi_{i+1}|=0.2, \quad {\phi_0=0}, \quad {\phi_1=5.08}
\label{eq:parameters_step2}
\end{split}
\end{equation}
{and} for initial condition we have $\phi_{ic}=6$. Hence, the prediction of spectral index and tensor-to-scalar ratio is given as follows:
\begin{equation}
n_s=0.96839,\quad r=0.00354 \, \, 
\end{equation}
which are in complete accordance with the observable constraints of inflation \cite{Planck:2018jri}.  We note here that  the number of efolds we obtain for this set of parameters  is $N_{tot}-N_{CMB} \approx 50$, where $N_{CMB}$ denotes the number of efolds at the pivot scale.  

For this enhancement,  the mass of PBHs is about $10^{-1} M_{\odot}$,  with the formalism of this evaluation presented in Appendix \ref{app:b}.  It is noteworthy  to mention here that recently  a few intriguing PBHs in low mass ratio binaries have been reported in GW observations \cite{LIGOScientific:2022hai,Phukon:2021cus}. Therefore, the predicted range of  masses of PBHs from this step-like mechanism can be related to the aforementioned searches, but then it needs further investigation
.

\begin{figure}[H]
\includegraphics[width=10.5 cm]{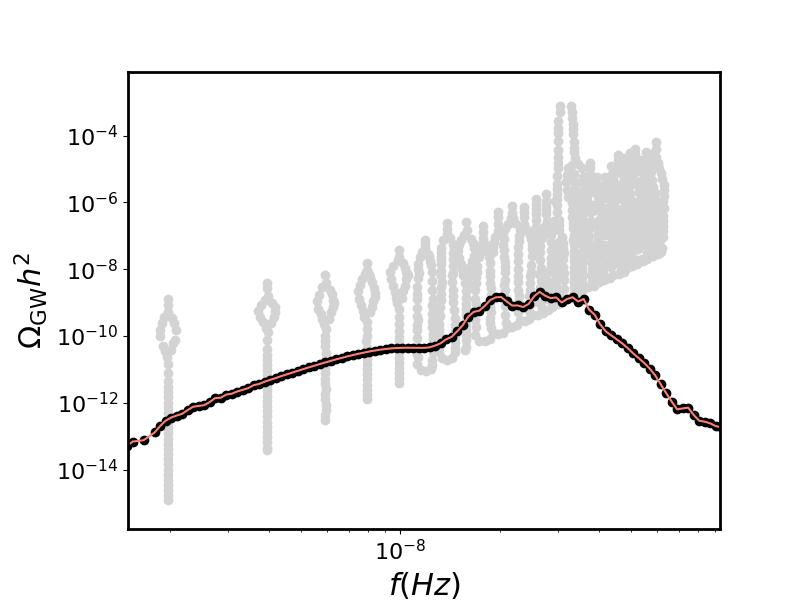}
\caption{The energy density of GWs  for the  potential given in Equation~(\ref{eq:steppot}), with the choice of parameters given in (\ref{eq:parameters_step2}).
\label{fig6}}
\end{figure}

In Figure~\ref{fig6}, we depict the energy density of GWs, denoted as $\Omega_{GW}$, using the specific parameters discussed in this section. The calculation of $\Omega_{GW}$ is detailed in Appendix \ref{app:c}. The gray dots on the plot represent the detection sensitivity of the NANOGrav collaboration.  Therefore, this mechanism has the interesting characteristic that it can be imprinted in the GW signals detected by NANOGrav \cite{NANOGrav:2023gor, NANOGrav:2023hde, NANOGrav:2023hvm}.
{It is important to note that the parameter values examined in Figure~\ref{fig6} do not cover  the energy densities of GWs observed  at larger frequencies, such as those from  LVK collaboration \cite{Abbott:2016blz, Abbott:2016nmj, Abbott:2017vtc, Abbott:2017gyy, Abbott:2017oio,dwyer2015gravitational} or those expected from future space-based experiments like LISA \cite{amaro2017laser} and DECIGO \cite{Seto:2001qf} or ground-based observatories such as ET \cite{Maggiore:2019uih}. However, adjusting the parameter scans may yield energy densities that can explain these larger-scale signals.  For a detailed review about the induced GWs and the frequency regions of future and present experiments, see Ref. \cite{Domenech:2021ztg}.  }

{Finally}, we observe that, in addition to this enhancement, {depicted in Figure~\ref{fig5},} there are also  additional oscillations that could be imprinted  {in the energy densities of the induced GWs shown in Figure~\ref{fig6}}. {These oscillations can} potentially lead to a characteristic \mbox{signal~\cite{Fumagalli:2020nvq,Mavromatos:2022yql}.} Moreover,  we need to mention here that in this work we neglect the possible effects
of non-Gaussianities \cite{Cai:2018dig,Perna:2024ehx,Liu:2023ymk,Namjoo:2012aa}. Consequently, this mechanism needs further investigation and should be the focus of a future study.

\section{Conclusions}
This study has provided an extensive evaluation of fine-tuning requirements across various inflationary models for the production of PBHs. We have demonstrated that while single-field models with inflection points require significant fine-tuning to align with CMB observations and produce PBHs, alternative models with step-like features in the scalar potential present a more favourable scenario. These models significantly reduce the need for fine-tuning and are capable of explaining the potential low-frequency GW signals detectable by PTA collaborations.

Further exploration into multi-field models reveals that the addition of extra fields can considerably alleviate the fine-tuning demands. Specifically, models incorporating non-canonical kinetic terms and hybrid inflationary models with a waterfall trajectory have shown promising results, reducing the fine-tuning requirements by orders of magnitude compared to  single-field models.

Moreover, the implementation of a spectator field offers a novel approach to circumvent the fine-tuning issue entirely. These models allow for the natural stochastic variations in field values across different Hubble patches, which can lead to the significant curvature fluctuations necessary for PBH production without stringent constraints on the inflationary potential's parameters.


{ To sum up, we examined several inflationary models for the production of PBHs and classified them based on the degree of fine-tuning required for explaining their copious production. We provided a comparative analysis of single- and multi-field inflationary mechanisms, highlighting the advantages of multi-field approaches in reducing the need for fine-tuning.   These insights help expand the theoretical framework connecting inflation, PBHs, and GWs. The novelty of this study lies in its contribution to guiding future research toward models that offer a more natural explanation for the formation of PBHs, thus paving the way for more grounded investigations in this area. }

\funding{The work of I.S. from the Belgian Francqui Foundation through a Francqui Start-up Grant, as well as the Belgian Fund for Scientific Research through an MIS grant and an IISN grant.}

\dataavailability{{No new data were created or analyzed in this study. Data sharing is not applicable to this article.}} 

\conflictsofinterest{The authors declare no conflicts of interest.} 

\label{sec:con}
\appendixtitles{no} 
\appendixstart
\appendix
\section[\appendixname~\thesection]{}
\label{app:a}
In this Appendix \ref{app:a}, we summarize the evaluation of the power spectrum that is needed in our analysis. {Additionally, we outline several noteworthy characteristics of the models explored in this study. }

In this framework, the field equations for ${\varphi^i}$, expressed in terms of efold time (assuming ${M_P=1}$), are provided by the following equation:
\begin{equation}
\ddot{\varphi}^c + \Gamma_{ab}^c \dot{\varphi}^a \dot{\varphi}^b + \left( 3 - \frac{1}{2} \dot{\sigma^2} \right) \dot{\varphi}^c + \left( 3 - \frac{1}{2} \dot{\sigma^2} \right) \frac{V^c}{V} = 0
\end{equation}
where the dots indicate derivatives with respect to efold time, and the indices ${i = a,b,c}$ denote derivatives relative to the fields. The Christoffel symbols $\Gamma^{c}_{ab}$ are defined  as:
\begin{equation}
\Gamma^{c}_{ab} = \frac{1}{2} G^{cd} (G_{da,b} + G_{db,a} - G_{ab,d}),
\end{equation}
where ${G_{ij}}$ represents the field metric. The velocity field $\dot\sigma$ is calculated as:
\begin{equation}
\dot{\sigma}^2 = G_{ab} \dot{{\varphi}^a} \dot{{\varphi}^b}.
\end{equation}

The Hubble parameter is defined by:
\begin{equation}
H^2 = \frac{V}{3 - \frac{1}{2} \dot{\sigma^2}}
\end{equation}
and the slow-roll parameter $\varepsilon_1$ is determined from:
\begin{equation}
\varepsilon_1 = \frac{1}{2} \dot{\sigma^2}.
\label{eq:epsilon_slowroll}
\end{equation}

The perturbation equations for the fields ${\delta\varphi}$ are given by:
\begin{equation}
\begin{split}
&\delta \ddot{\varphi}^c + (3 - \varepsilon_1) \delta \dot{\varphi}^c + 2 \Gamma^c_{ab} \dot{\varphi}^a \delta \dot{\varphi}^b + \left( \Gamma^{c}_{ab,d} \dot{\varphi}^a \dot{\varphi}^b + \frac{V^{c}_{d}}{H^2} - G^{ca} G_{ab,d} \frac{V^b}{H^2} \right) \delta \varphi^d \\
& + \frac{k^2}{a^2 H^2} \delta \varphi^c = 4 \dot{\Psi} \dot{\varphi} - 2 \Psi \frac{V^c}{H^2}
\end{split}
\label{eq:perturbation_of_field}
\end{equation}
and the equation for the Bardeen potential $\Psi$ is:
\begin{equation}
\ddot{\Psi} + (7 - \varepsilon_1) \dot{\Psi} + \left( 2 \frac{V}{H^2} + \frac{k^2}{a^2 H^2} \right) \Psi = -\frac{V_c}{H^2} \delta \varphi^c.
\end{equation}

We assume the initial condition of a Bunch--Davies vacuum. The power spectrum is described by:
\begin{equation}
P_R = \frac{k^3}{(2 \pi)^2} (R_i^2)
\label{eq:pr_general}
\end{equation}
where:
\begin{equation}
R_i = \Psi + H \sum_i^{N_\text{fields}} \frac{\dot{\varphi_i} \delta \varphi_i }{\dot{\varphi_i}^2}.
\end{equation}

{In the case of single-field inflation $\varphi^1=\phi$,  the equation of the curvature perturbations takes the following form \cite{Kefala:2020xsx}:}

\begin{equation}
{R''}_k+f(N) R'_k +\frac{k^2}{H^2}e^{-2N}R_k=0
\end{equation}
{ where  }
\begin{equation}
f(N)= 3+2 \frac{\phi''}{\phi'}-\frac{\phi'^2}{2}\, .
\label{eq:app:frict}
\end{equation}

{This function acts as a friction term.}

\begin{figure}[H]
\includegraphics[width=6.5 cm,height=4cm]{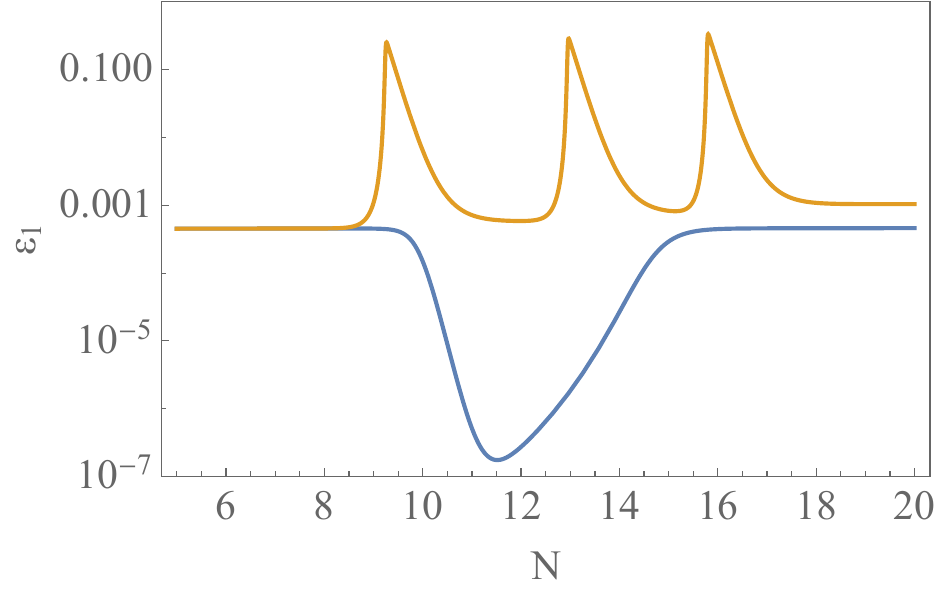}
\includegraphics[width=6.5 cm,height=4.2cm]{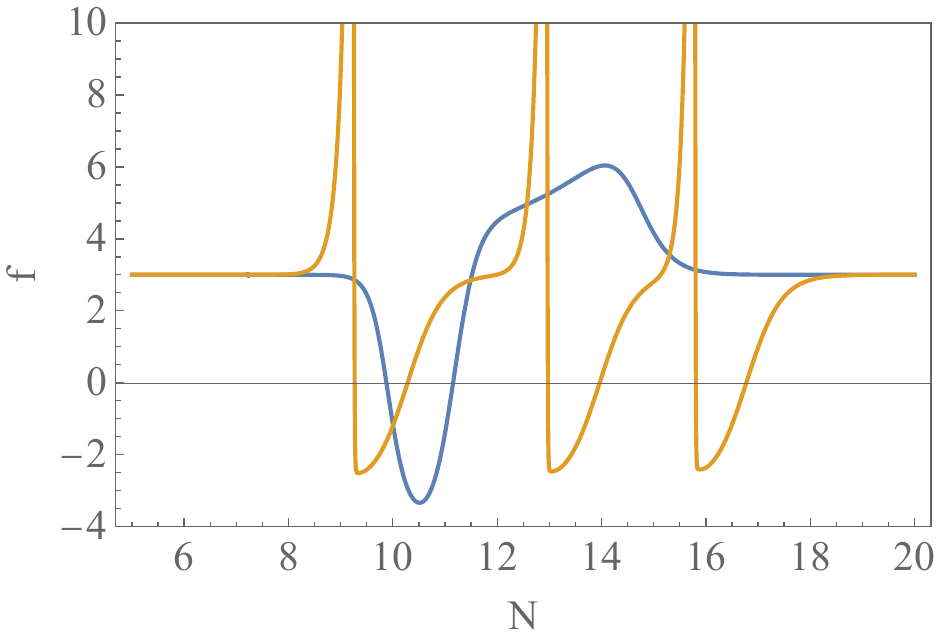}
\caption{The different behavior of the slow-roll  parameter $\varepsilon_1$ (\textbf{left panel}) and friction term $f$ (\textbf{right panel})  for the mechanism of an inflection point (blue line) and the one with the step-like potential (orange line). For blue line: $i=1$, $A_1 = 0.000605$, $c=100$, $B = -0.03$. For orange  line: $i = 3$, $A_1=0.1$, $c = 100$ $B = 0.03$, $\varphi_1 = 0.4$, and $\varphi2 = 0.8$.  For initial condition, $\phi_{ic} = -0.3$. 
\label{fig101}}
\end{figure}

{  The potential in Equation~(\ref{eq:steppote}) can interplay between the mechanism of the inflection point and that of the step-like potential
.}
{In Figure~\ref{fig101}, we illustrate the behavior of the slow-roll parameter $\varepsilon_1$,  Equation~(\ref{eq:epsilon_slowroll}), and the $f$ from Equation~(\ref{eq:app:frict}) for two different scenarios: an inflection point (blue curve) and a step-like potential (orange line), using the potential specified in Equation~(\ref{eq:steppote}). The parameters for each scenario are provided in the figure. This plot highlights the distinct dynamics between these two mechanisms; in the inflection point scenario, the velocity of the field significantly decreases, whereas in the step-like potential, the field's velocity exhibits both increases and decreases. Moreover, the function $f$   behaves differently in the cases of the inflection point and step-like potential
. Hence, through this figure we can notice that the previously mentioned mechanisms behave differently. }

\section[\appendixname~\thesection]{}
\label{app:b}

{In Appendix \ref{app:b}, we briefly discuss the methodology  to evaluate the fractional abundance of PBHs. First, we outline the theoretical framework and computational approaches used to estimate the PBH abundance. Subsequently, we present  the fractional abundances calculated for the various inflationary models discussed within this work. Finally, we discuss   these results.} 

 The {fractional} abundance of PBH is given by the integral:
\begin{equation}
f_{PBH} = \int d\ln M_ {PBH} \frac{\Omega_ {PBH}}{\Omega_ {DM}}
\end{equation}
\noindent
where
\begin{equation}
\label{44}
\frac{\Omega_ {PBH}}{\Omega_ {DM}}= \frac{\beta(M_\text{PBH}(k))}{8 \times 10^{-16}} \left(\frac{\gamma}{0.2}\right)^{3/2} \left(\frac{g(T_f)}{106.75}\right)^{-1/4}\left(\frac{M_\text{PBH}(k)}{10^{-18} \textsl{g}}\right)^{-1/2}.
\end{equation}

Here, $\gamma$ is a factor that depends on gravitation collapse,
and we choose $\gamma =0.2$~\cite{Carr:1975qj}.
 $T_f$   denotes  the temperature  of PBH formation, and $M_\text{PBH}$ is the mass of PBHs.    $g(T_f)$  are the effective degrees of freedom during this formation, and counting only the 
 SM particles we set $g(T_f)=106.75$. 
 
  {The mass fraction of PBHs at formation, denoted by $\beta(M_\text{PBH})$, can be estimated using either the Press–Schechter (PS) approach~\cite{Press74} or Peak Theory~\cite{Bardeen:1985tr}. In this section, we focus on the PS approach. }
{Hence,  the mass fraction of   $\beta(M_\text{PBH})$ is given as:}
\begin{equation}
\label{42}
\beta(M_\text{PBH})= \frac{1}{\sqrt{2 \pi \sigma ^2 (M_\text{PBH})}} \int^{\infty}_{\delta_c} d\delta \,  \exp \left(  -\frac{\delta ^2}{2 \sigma^2(M_\text{PBH}) } \right) \, ,
\end{equation}
where {the overdensity $\delta$ is above a certain threshold of collapse $\delta_c$. The variance of curvature perturbation is given} : 
\begin{equation}
\label{40}
\sigma^2 \left( M_\text{PBH}(k)  \right)= \frac{16}{81}  \int \frac{dk' }{k'} \left(\frac{k'}{k}\right)^4 P_\mathcal{R}(k') W^2\left(\frac{k'}{k}\right),
\end{equation}
where W is a window function. 

The mass as a function of the comoving wavenumber $k$ \cite{Ballesteros:2017fsr}:
\begin{equation}
\label{43}
M_{PBH}=10^{18}  \left(\frac{\gamma}{0.2}\right)  \left(\frac{g(T_f)}{106.75}\right)^{-1/6} \left(\frac{k}{7 \times 10^{13} Mpc^{-1}  }\right)^{-2} \textsl{g} \, .
\end{equation}

\begin{figure}[H]\vspace{-16pt}
\includegraphics[width=10.5 cm]{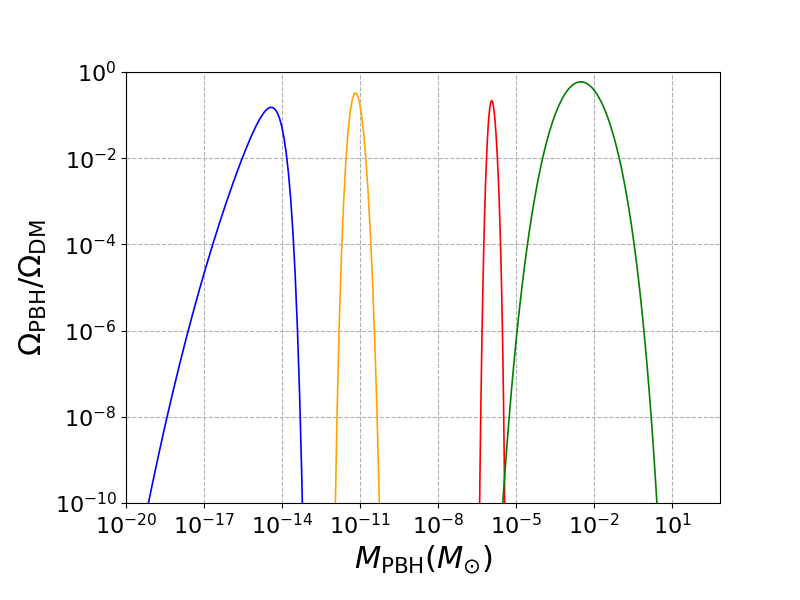}
\caption{The fractional abundances of PBHs for the four different mechanisms: inflection point (blue), step-like feauture potential (red), two field with a non-canonical kinetic term (orange), and hybrid model (green).
\label{fig7}}
\end{figure} 


{In Figure~\ref{fig7}, we illustrate the fractional abundances of PBHs for various mechanisms: an inflection point, step-like features in the potential, a two-field model with non-canonical kinetic terms in the Lagrangian, and a two-field model with a waterfall trajectory, each represented by distinct colors as detailed in the caption. We use the parameter sets presented in Figures \ref{fig1}--\ref{fig4}, which lead to an enhancement of the scalar power spectrum of at least $\rm{10^{-2}}$.  We can notice that models with step-like potential can lead to narrower mass ranges. The models shown in Figure~\ref{fig7} can account for a wide range of PBH masses. However, it is important to emphasize that different parameter choices can yield, varying fractional abundances; therefore, this plot should not be solely relied upon for comparing these mechanisms. Furthermore, the fractional abundances can exceed constraints, such as those given in Refs. \cite{Capela:2013yf,Niikura:2017zjd}. As mentioned, the proper adjustment of parameters for each model is required, but this is beyond the scope of this paper. }

{ For the evaluation of fractional abundance depicted in Figure~\ref{fig7}, we adopt critical threshold values ranging from $\delta_c=0.4$ to $0.6$ as referenced in Ref. \cite{Musco:2020jjb,Escriva:2020tak,Escriva:2022duf}. It is crucial to compute the threshold value from the respective power spectrum for each scenario, as outlined in \cite{Musco:2020jjb,Escriva:2020tak,Escriva:2022duf,Stamou:2023vxu}. \textls[-15]{Nonetheless, by appropriately adjusting the model parameters, one can achieve the desired fractional abundances. The choice of window function, specifically a Gaussian window function as applied in this analysis (see Equation~(\ref{40})), significantly influences the threshold calculation. Although the choice of window function introduces corrections in the threshold computation, these impacts are minimized when the same window function is consistently used across evaluations of PBH abundance~\cite{Kalaja:2019uju,Young:2019osy, Tokeshi:2020tjq, Musco:2020jjb}.}}

\section[\appendixname~\thesection]{}
\label{app:c}
In this Appendix, we briefly discuss how to evaluate the energy density of GWs from the scalar power spectrum.  {The stochastic background of GWs can be generated through the coupling of the scalar and tensor modes
}~\cite{Espinosa:2018eve,Kohri:2018awv,Mollerach:2003nq,Maggiore:1999vm,Baumann:2007zm}.

The energy density of the GWs  can be evaluated from the following integral: 
\begin{equation}
{\Omega_{GW}(k)}=\frac{\Omega_r}{36} \int^{\frac{1}{\sqrt{3}}}_{0}\mathrm{d} d
\int ^{\infty}_{\frac{1}{\sqrt{3}}}\mathrm{d} s \left[  \frac{(s^2-1/3)(d^2-1/3)}{s^2+d^2}\right]^2\, 
P_{R}(kx) \,  P_{R}(ky) \,(I_c^2+I_s^2) \, ,
\label{eq4.1}
\end{equation}
where the radiation density  $\Omega_r \approx 5.4 \times 10^{-5}$. The variables $x$ and $y$ are
\begin{equation}
x= \frac{\sqrt{3}}{2}(s+d), \quad  y=\frac{\sqrt{3}}{2}(s-d).
\label{eq4.2}
\end{equation}

The functions $I_c$ and $I_s$ are given by the expressions 
\begin{gather}
I_c=-36 \pi \frac{(s^2+d^2-2)^2}{(s^2-d^2)^3}\Theta(s-1)\\
I_s=-36 \frac{(s^2+d^2-2)^2}{(s^2-d^2)^2}\left [ \frac{(s^2+d^2-2)}{(s^2-d^2)} \log\left| \frac{d^2-1}{s^2-1} \right| +2\right].
\label{eq4.3}
\end{gather}

Using the fact that  
\begin{center}
$1 \, {\rm Mpc}^{-1}= 0.97154\times \, 10^{-14} \,  {\rm s}^{-1}$ and $k=2\pi \rm{f}$
\end{center}

{we} can evaluate the energy density of the power spectrum as a function of the frequency $\rm{f}$.

\begin{adjustwidth}{-\extralength}{0cm}

\reftitle{References}


\PublishersNote{}
\end{adjustwidth}
\end{document}